\documentclass[aps,prb,twocolumn,superscriptaddress,showkeys,amsmath,amssymb]{revtex4-1}
\usepackage{graphicx}
\usepackage{dcolumn}
\newcolumntype{.}{D{.}{.}{-1}}
\AtBeginDocument{\usepackage{booktabs}}
\usepackage{hyperref}

\begin{document} 

\title[Surface Dynamics of Bi$_2$Te$_3$(111)]{Nanoscale Surface Dynamics of Bi$_2$Te$_3$(111): Observation of a Prominent\\ Surface Acoustic Wave and the Role of van der Waals Interactions}

\author{Anton Tamt\"{o}gl}
\email{tamtoegl@gmail.com}
\affiliation{Cavendish Laboratory, J. J. Thompson Avenue, Cambridge CB3 0HE, United Kingdom}
\affiliation{Institute of Experimental Physics, Graz University of Technology, 8010 Graz, Austria}
\author{Davide Campi}
\affiliation{Theory and Simulation of Materials (THEOS) and National Centre for Computational Design and Discovery of Novel Materials (MARVEL), \'{E}cole Polytechnique F\'{e}d\'{e}rale de Lausanne, Lausanne CH-1015, Switzerland}
\author{Martin Bremholm}
\author{Ellen M. J. Hedegaard}
\author{Bo B. Iversen}
\affiliation{Center for Materials Crystallography, Department of Chemistry and iNANO, Aarhus University, 8000 Aarhus, Denmark}
\author{Marco Bianchi}
\author{Philip Hofmann}
\affiliation{Department of Physics and Astronomy, Interdisciplinary Nanoscience Center (iNANO), Aarhus University, 8000 Aarhus, Denmark}
\author{Nicola Marzari}
\affiliation{Theory and Simulation of Materials (THEOS) and National Centre for Computational Design and Discovery of Novel Materials (MARVEL), \'{E}cole Polytechnique F\'{e}d\'{e}rale de Lausanne, Lausanne CH-1015, Switzerland}
\author{Giorgio Benedek}
\affiliation{Dipartimento di Scienza dei Materiali, Universit\'a di Milano-Bicocca, Via R. Cozzi 53, 20125 Milano, Italy}
\affiliation{Donostia international Physics Center (DIPC), University of the Basque Country (EHU-UPV), Donostia, San Sebastian, Spain}
\author{John Ellis}
\author{William Allison}
\affiliation{Cavendish Laboratory, J. J. Thompson Avenue, Cambridge CB3 0HE, United Kingdom}

\begin{abstract}
We present a combined experimental and theoretical study of the surface vibrational modes of the topological insulator Bi$_2$Te$_3$. Using high-resolution helium-3 spin-echo spectroscopy we are able to resolve the acoustic phonon modes of Bi$_2$Te$_3$(111). The low energy region of the lattice vibrations is mainly dominated by the Rayleigh mode which has been claimed to be absent in previous experimental studies. The appearance of the Rayleigh mode is consistent with previous bulk lattice dynamics studies as well as theoretical predictions of the surface phonon modes. Density functional perturbation theory calculations including van der Waals corrections are in excellent agreement with the experimental data. Comparison of the experimental results with theoretically obtained values for films with a thickness of several layers further demonstrate, that for an accurate theoretical description of three-dimensional topological insulators with their layered structure the inclusion of van der Waals corrections is essential. The presence of a prominent surface acoustic wave and the contribution of van der Waals bonding to the lattice dynamics may hold important implications for the thermoelectric properties of thin-film and nanoscale devices. 
\end{abstract}


\maketitle 

\section{Introduction}
Bi$_2$Te$_3$ is one of the most studied topological insulators (TI)\cite{Chen2009}, a class of materials which exhibit protected metallic surface states and an insulating bulk electronic structure\cite{Hasan2010,Moore2009}. The interaction of surface phonons with electrons on Bi$_2$Te$_3$ has been mainly focused on determinations of the electron-phonon (e-ph) coupling constant $\lambda$\cite{Tamtogl2017,Sobota2014,Howard2014,Kondo2013}. Since scattering channels for the surface state electrons may impose constraints for potential applications such as surface-dominated transport, $\lambda$ is a convenient parameter to characterise the e-ph interaction strength\cite{Barreto2014}.\\
However, very limited experimental data exists for the surface phonon dispersion of Bi$_2$Te$_3$(111) as well as for the region of acoustic phonon modes on topological insulator surfaces in general. Information on the surface phonon dispersion is also essential to fully understand the thermoelectric properties of Bi$_2$Te$_3$ thin films and nanoscale devices\cite{Minnich2009,Liang2016,Rojo2017} and the low lattice thermal conductivity of Bi$_2$Te$_3$, one of the most studied and efficient thermoelectric materials\cite{Rittweger2014,Snyder2008}. Previously, the excellent thermoelectric performance of Bi$_2$Te$_3$ has been attributed to the details of the electronic structure and a low lattice thermal conductivity similar to ordinary glass\cite{Snyder2008,Shi2015}. While the vibrational properties of Raman active modes have been studied\cite{Chis2012,Wang2013}, a systematic experimental investigation of the acoustic phonons in these materials is still missing.\\ 
Here we use the helium spin-echo technique which is capable of measuring surface phonon spectra with very high resolution\cite{Kole2010}. He atom beams with energies of typically 10 meV are perfectly suited to probe all kind of surfaces in an inert, completely nondestructive manner\cite{Farias1998}. In the context of TI surfaces, helium atom scattering (HAS) has also the advantage that the samples are not exposed to any intense ultraviolet illumination which has been reported to trigger energetic shifts of the electronic bands near the surface of a TI crystal after cleaving.\cite{Frantzeskakis2017}. Since He atoms at thermal energies are scattered directly by the surface charge density corrugation\cite{Mayrhofer2013}, inelastic scattering results from coupling to phonon-induced charge density oscillations (CDOs). Therefore, inelastic HAS can also be used to infer information about the corresponding e-ph coupling strength\cite{Manson2016,Tamtogl2017}.\\
While previous experimental studies claimed that the Rayleigh wave (RW) is absent on Bi$_2$Te$_3$(111)\cite{Howard2013}, theoretical studies showed that the e-ph interaction cannot account for this absence\cite{Parente2013,Heid2017}. Moreover, the low energy region of surface acoustic waves ($\approx 0-5$ meV) has not been subject to an experimental study up to now and we show, that the RW is not only present but also the dominant feature in the inelastic scattering spectra in this energy region.\\
Furthermore, we compare the experimentally determined surface phonon dispersion with density functional perturbation theory (DFPT) calculations detailing the role of van der Waals (vdW) interactions and spin-orbit coupling (SOC). As demonstrated recently, many-body vdW interactions can substantially affect vibrational properties leading to the appearance of low-frequency phonon modes which may even favour the thermal stability of a certain crystal structure\cite{Reilly2014,Folpini2017}. The inclusion of vdW interactions becomes increasingly important for layered TIs\cite{Bjorkmann2012} and two-dimensional materials such as e.g. hexagonal boron nitride\cite{Vuong2017,Cusco2018} or for a correct description of hydrogenated graphene\cite{Birowska2017,Bahn2017}.

\section{Experimental and Computational Details}
\begin{figure}[htb]
\centering
\includegraphics[width=0.42\textwidth]{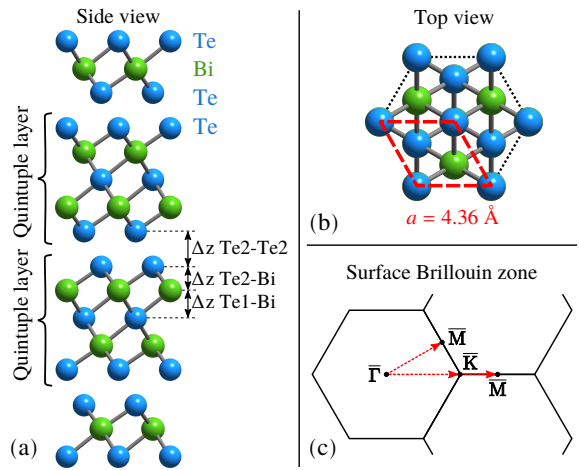}
\caption{Crystal structure of Bi$_2$Te$_3$. (a) Side view of the conventional hexagonal unit cell containing three quintuple layers of which each one is terminated by a Te layer. (b) Top view of the (111) surface of Bi$_2$Te$_3$ where the red rhombus highlights the hexagonal surface unit cell. (c) Surface Brillouin zone with the two scanning directions.}
\label{fig:struct}
\end{figure}

The reported measurements were performed on the Cambridge helium-3 spin-echo apparatus which generates a nearly monochromatic beam of $^3$He that is scattered off the sample surface in a fixed 44.4$^{\circ}$ source-target-detector geometry. For a detailed description of the apparatus please refer to\cite{Alexandrowicz2007,Jardine2009}.\\
The rhombohedral crystal structure of Bi$_2$Te$_3$ consists of quintuple layers (QLs) which are connected by weak vdW forces\cite{Michiardi2014}. The hexagonal unit cell of the Bi$_2$Te$_3$ crystal is shown in \autoref{fig:struct} which consists of 3 QLs. Each quintuple layer (QL) is terminated by Te atoms, giving rise to the (111) cleavage plane that exhibits a hexagonal structure ($a=4.36~\mathrm{\AA}$\cite{Tamtogl2017}, see \autoref{fig:struct}(b)). The Bi$_2$Te$_3$ single crystal used in the study was attached onto a sample cartridge using electrically and thermally conductive epoxy. The sample cartridge was then inserted into the scattering chamber using a load-lock system\cite{Tamtogl2016a} and cleaved \textit{in-situ}. The sample can be heated using a radiative heating filament on the backside of the crystal or cooled down to $110~\mbox{K}$ via a thermal connection to a liquid nitrogen reservoir. The sample temperature was measured using a chromel-alumel thermocouple.\\
All scattering spectra were taken with the crystal cooled down to 115 K. The incident He beam energy was set to 8 meV with the exception of a few spectra that were measured at an incident beam energy of 12 meV.

\subsection{Computational Details}
\label{sec:CompDetails}
The dynamical properties of Bi$_2$Te$_3$ were studied by means of density functional perturbation theory (DFPT)\cite{Baroni2001} as implemented in the Quantum-ESPRESSO package\cite{QMEspresso}. The surface phonon dispersion of Bi$_2$Te$_3$(111) was calculated using a slab consisting of 3 QLs separated from its periodic replica by 30 \AA\ of vacuum. For an accurate calculation of the surface lattice vibrations, in principle both SOC and a proper treatment of vdW corrections are necessary. The presence of heavy elements in Bi$_2$Te$_3$ requires the inclusion of SOC while at the same time the inclusion of vdW interactions is necessary for a correct description of the weak bonding between the individual quintuple  layers\cite{Bjorkmann2012}. To asses the importance of SOC and vdW interactions we performed the calculations using a standard Perdew-Burke-Ernzerhof (PBE) functional\cite{pbe} and two flavours of vdW correction, a semiempirical DFT-D3 correction\cite{Grimme2010} and a nonlocal vdW-DF2-C09 functional\cite{vdwdf,vdwdf2,vdwdf3}. The first two were performed with the inclusion of SOC while the latter only without SOC due to the limitations of the current implementation in Quantum-ESPRESSO. Hermann \emph{et al.}\cite{Hermann2017} provide a recent discussion about the advantages and limitations of existing vdW-inclusive methods.\\
Effects of SOC were treated self-consistently with fully relativistic pseudopotentials and the formalism for noncollinear spin magnetisation. We used scalar relativistic and fully relativistic norm conserving pseudopotentials from the PseudoDojo library with an energy cutoff of 80 Ry. The surface Brillouin zone (SBZ) was sampled using a $9\times 9\times 1$ uniform $k$-point grid\cite{mp}.\\
Due to the topologically non-trivial nature of Bi$_2$Te$_3$ the inclusion of SOC leads to the formation of Dirac cones in the surface electronic structure. To deal with this metallic nature, a Gaussian smearing of 0.01 Ry was introduced in the occupation of states. The topologically protected Dirac cone forms a small Fermi circle around the $\Gamma$-point which cannot be properly described with a coarse mesh and a large smearing resulting in the impossibility to capture subtle effects, such as the proposed Kohn anomaly\cite{Howard2013} with a standard calculation. For this reason we also performed calculations using a non-uniform $k$-point grid with a density equal to a $90\times 90 \times 1$ grid around $\Gamma$ and a $9\times 9 \times 1$ grid far away from it with a reduced smearing of $0.001~$Ry. Due to the high computational cost these calculations were performed only at the $\Gamma$-point and at a single $q$-point corresponding to the nesting vector of the Fermi surface $q_{Nest}=2k_{F}=0.28~\mathrm{\AA}^{-1}$ where the Kohn anomaly is expected. To reproduce the experimental nesting vector, determined by the self-doping of the sample, a fraction of $0.016$ electrons was added to the simulation.

\section{Results and discussion}
\begin{figure}[htb]
\centering
\includegraphics[width=0.48\textwidth]{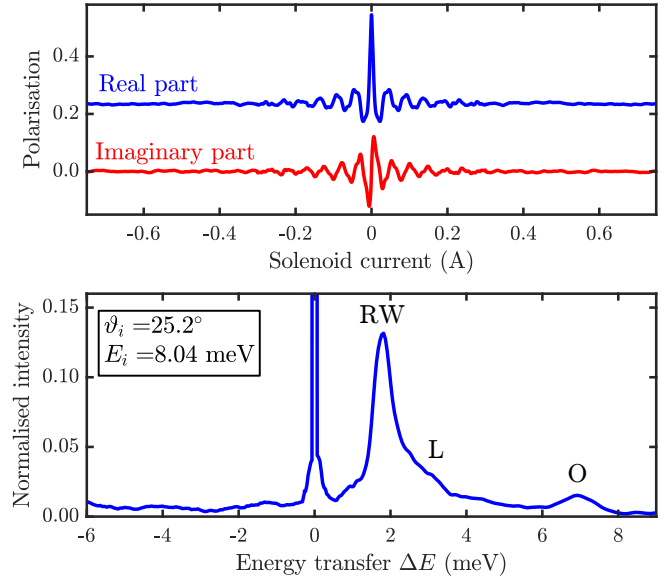}
\caption{A typical $^3$He spin-echo  measurement performed along the $\overline{\mathrm{\Gamma M}}$ azimuth with the crystal at 115 K. The upper panel shows the real and imaginary components of the complex polarisation. The oscillations are due to phonon scattering events which becomes evident after Fourier transforming and converting to the energy transfer scale (lower panel). The peak at $\Delta E = 0$ is due to diffuse elastic scattering, and the phonon modes are now visible as finite energy transfer peaks with the Rayleigh wave (RW), the longitudinal resonance (L) and an optical mode (O). The most dominant phonon peak corresponds to the Rayleigh mode and appears at $\Delta E \approx + 1.8 $ meV in the spectrum.}
\label{fig:ExampleISF}
\end{figure}

\begin{figure}[htb]
\centering
\includegraphics[width=0.48\textwidth]{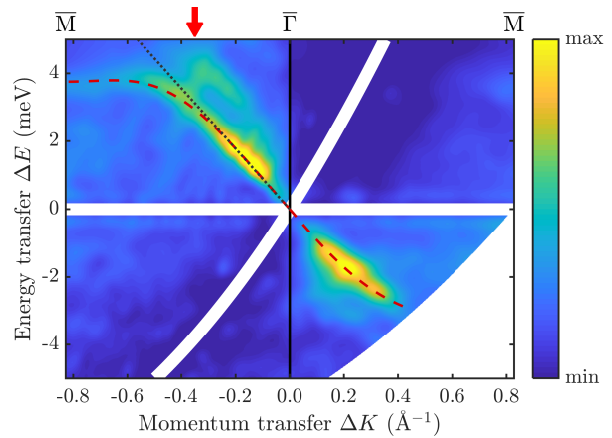}
\caption{(Colour online) Collection of several experimental spectra along  $\overline{\mathrm{\Gamma M}}$ with an incident beam energy of 8 meV. The intensity is shown as a function of the energy and the parallel momentum transfer. The region around $\Delta E =0$ is not shown due to the large intensity of the elastic peak compared to the phonon modes. The red dashed line is plotted as a guide to the eye to show the dispersion of the Rayleigh mode. The black dotted line represents the group velocity of the Rayleigh mode.}
\label{fig:3D_phonon}
\end{figure}

Surface phonon energies were obtained by performing spin-echo measurements over a wide range of spin-echo times (including both real and imaginary components of the beam polarisation). Therefore the solenoid current was varied from $-1$ to $+1$ A with 2049 equally spaced points (top panel in \autoref{fig:ExampleISF}). Oscillations in the polarisation correspond to surface atoms vibrating with a characteristic period, i.e. to a particular surface phonon mode. Hence Fourier transforming the data to the wavelength domain and converting to the energy scale gives rise to spectra which are analogous to time-of-flight (TOF) spectra, with energy loss and gain peaks for the creation and annihilation of a phonon, respectively\cite{Jardine2009,Kole2010} (lower panel in \autoref{fig:ExampleISF}). The phonon energy $\Delta E =  \hbar \omega$ is then given via: 
\begin{equation}
\Delta E =  \hbar\omega= \frac{h^2}{2m}\left(\frac{1}{\lambda_f^2}-\frac{1}{\lambda_i^2}\right). 
\label{eq:freq_lambda}
\end{equation}
where $\lambda_i$ and $\lambda_f$ are the incoming and outgoing wavelength of the He beam and $m$ is the $^3$He mass.\\
To determine the entire phonon dispersion curve up to the Brillouin zone boundary a series of spin-echo spectra at incident angles between the first-order diffraction peaks was measured. The phonon dispersion was then obtained by calculating the parallel momentum transfer $|\Delta K|$ for each extracted phonon energy $\Delta E$ from the conservation laws of energy and parallel momentum, providing the so-called scancurve for planar scattering \cite{Kole2010,Benedek1975}: 
\begin{equation}
\frac{\Delta E}{E_i} + 1 = \frac{\sin^2 \vartheta_i}{\sin^2 \vartheta_f} \left( 1 + \frac{\Delta K}{K_i} \right)^2 
\label{eq:scancurve}
\end{equation} 
where $E_i$ is the energy of the incident beam and $K_i$ is the parallel component of the incident wavevector. Here, $\vartheta_i$ and $\vartheta_f$ are the incident and final angle with respect to the surface normal, respectively. The phonon parallel momentum is then given by $|\Delta K|+G$, where the surface reciprocal lattice vector $G$ in the scattering plane, needed to bring $Q$ into the first Brillouin zone, accounts for \emph{umklapp} processes.\\ 
\autoref{fig:3D_phonon} shows a collection of 40 inelastic scattering spectra measured along $\overline{\mathrm{\Gamma M}}$ with an incident beam energy of 8 meV. Each spectrum was converted to energy $\Delta E$ and parallel momentum $\Delta K$ using \eqref{eq:freq_lambda} and \eqref{eq:scancurve} and the intensity is plotted as a function of $(\Delta K, \Delta E)$. The region around $\Delta E =0$ is not shown due to the large intensity of the elastic peak compared to the phonon modes, as well as the region corresponding to $(\Delta K, \Delta E)$ very close to the specular condition ($\vartheta_i = \vartheta_f $). Most intensity is found along the dashed line and is associated with the RW, while the weaker intensity above the RW, corresponding to the shoulder labelled L in the lower panel of \autoref{fig:ExampleISF}, is attributed to the longitudinal resonance\cite{Benedek2010}.\\
Such a prominent RW could not be observed on TI surfaces in previous experiments and it was instead claimed that the RW mode is absent on Bi$_2$Te$_3$(111) in a study by Howard \emph{et al.}\cite{Howard2013}. However, the mentioned previous experimental study\cite{Howard2013} was rather concentrated on phonon events with higher energies and in the presence of a broad diffuse elastic peak, individual phonon peaks with small energy transfers (as the low-lying RW) are difficult to separate. Moreover, the quality of the crystal surface (as indicated by the diffraction spectra\cite{Tamtogl2017}) and the presence of defects such as steps will increase the height of the diffuse elastic peak. Finally, the energy resolution of the apparatus will add to the width of the peak, emphasising the advantage of the spin-echo technique in this context: In a standard spin-echo measurement, tuned for the quasi-elastic peak (around $\Delta E = 0$), the quasi-elastic measurement is only limited by the maximum Fourier time of the instrument\cite{Jones2016}. Note however, that under this measurement condition, the finite energy transfer peaks are also broadened by the projection of the feature in the wavelength intensity transfer matrix\cite{Jones2016,Kole2010}. It should also be mentioned that it is often difficult to directly compare intensities obtained via TOF measurements (as performed by Howard \emph{et al.}\cite{Howard2013}) with those from spin-echo measurements obtained after Fourier transforming the data.\\
Indeed the presence of the  RW is in line with the fact that the anisotropy of the elastic constants of Bi$_2$Te$_3$ is not strong at all\cite{Kullmann1990,Jenkins1972,Akgoz1972} and later theoretical studies about the electron-phonon interaction of Bi$_2$Te$_3$\cite{Parente2013,Giraud2012,Heid2017}. The bulk phonon properties of Bi$_2$Te$_3$ have been studied by Raman spectroscopy\cite{Richter1977,Kullmann1984,Wang2013,Chis2012} and an early neutron scattering study performed by Wagner \emph{et al.}\cite{Wagner1978}. Based on these experimental results together with lattice dynamical shell model calculations, Kullmann \emph{et al.}\cite{Kullmann1990} showed that in terms of the bulk lattice dynamics the typical properties of a three-dimensional crystal dominate: Kullmann \emph{et al.} concluded that the layered quintuple structure of Bi$_2$Te$_3$ gives rise to much less characteristic features of low dimensional crystal structures\cite{Kullmann1990} compared to two-dimensional materials such as graphene\cite{AlTaleb2016}.\\
The group velocity $v_g = \partial \omega / \partial K  $ of an acoustic phonon mode corresponds to the speed of sound along a certain crystal direction in the long wavelength limit (close to $\overline{\Gamma}$, see the black dotted line in \autoref{fig:3D_phonon}). Using the slope close to $\overline{\Gamma}$ we obtain for the speed of sound of the Rayleigh wave at 115 K and along both azimuthal directions:
\begin{align*}
\overline{\mathrm{\Gamma M}}: \; & v_{RW} = 1300~\mbox{m}/\mbox{s} \\
\overline{\mathrm{\Gamma K}}: \; & v_{RW} = 1600~\mbox{m}/\mbox{s}.
\end{align*}
In general the speed of surface acoustic waves is lower than the speed of the slowest bulk wave. The speed of sound in the bulk can be obtained from the elastic constants\cite{Farnell1970,Farnell1978} as determined from ultrasonic measurements at room temperature in the case of Bi$_2$Te$_3$\cite{Jenkins1972,Akgoz1972}. The values for shear vertical (SV) and shear horizontal (SH) polarisation along both azimuthal directions are:
\begin{align*}
\overline{\mathrm{\Gamma M}}: \; & v_{SV} \approx v_{SH} = 1740~\mbox{m}/\mbox{s} \\
\overline{\mathrm{\Gamma K}}: \; & v_{SV}  = 2290~\mbox{m}/\mbox{s},
\end{align*}
similar to the values found in later studies\cite{Hellmann2014,Chen2013a,Bessas2012}. The $v_{RW}/v_{SV}$ ratio for the surface modes as given by the two velocities along $\overline{\Gamma \mathrm{M}}$ and $\overline{\Gamma \mathrm{K}}$, is about 0.72. It is interesting to note that this value is similar to the ratio $v_{RW}/v_{SV} = 0.83$ which would be obtained for the $(001)$ surface of a cubic crystal with the same elastic constant ratios $C_{12} / C_{44} = 0.795$ and $C_{11} / C_{44} = 2.24$ of Bi$_2$Te$_3$ (see \cite{Farnell1970}). Moreover, $v_{RW}$ from the present study is somewhat smaller than that recently measured with Brillouin scattering in Bi$_2$Te$_3$ thin films grown on a stiffer substrate and with the RW penetration depths comparable to the film thickness\cite{Wiesner2017}.\\
Note also, that an important aspect of the surface dynamics of Bi$_2$Te$_3$ is that the above described RW is, technically, a pseudo-surface wave (PSW), because there exists a bulk acoustic branch which is even softer\cite{Wiesner2017}. From ultrasonic experiments\cite{Jenkins1972,Akgoz1972,Kullmann1990} it is known that an additional low lying bulk mode with shear horizontal (SH) polarisation exists (with a group velocity of about $1250~\mbox{m}/\mbox{s}$ along $\overline{\Gamma \mathrm{K}}$ as obtained from the elastic constants\cite{Jenkins1972,Akgoz1972,Huang2008}). In any direction where the sagittal plane (defined by the incident wave vector and the surface normal) has a mirror symmetry, the RW is orthogonal to the SH modes and therefore there is no mixing, and its localisation is ensured. However, any small deviation from the mirror-symmetry direction yields a coupling to bulk SH modes and the RW becomes a resonance, acquiring an oscillating factor on top of the exponential decay inside the bulk. Thus in experiments, where the angular acceptance angle cannot be zero, PSWs always have a resonance width and the peaks always show an intrinsic broadening with respect to ordinary localised RWs, due to the mixing with the SH modes.\\
Eventually, the comparison has to be made with surface dynamics calculations since bulk modes with a certain polarisation tend to give rise to surface modes with mixed polarisations upon projection onto the surface. In the following we will compare the measured surface phonon dispersion over the entire Brillouin zone with DFPT calculations. By inclusion of vdW interactions  we attempt to address the question whether dispersion corrections play an important role.
\subsection{DFPT dispersion curves and comparison with experiment}
\begin{figure}[htb]
\centering
\includegraphics[width=0.48\textwidth]{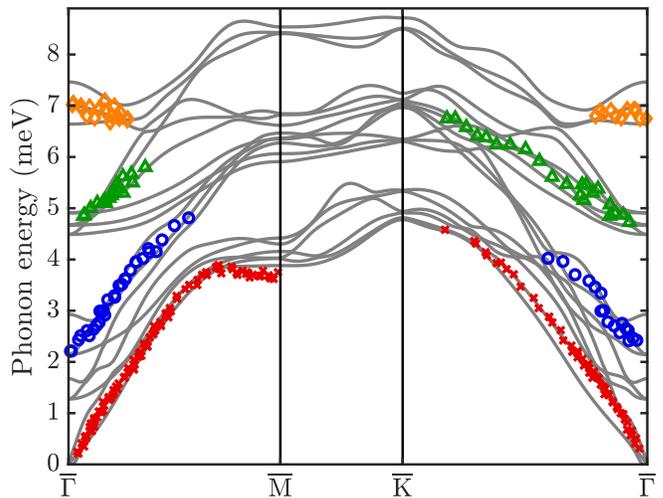}
\caption{Comparison of the measured phonon dispersion relation with DFPT calculations using the vdW-DF2-C09 functional for 3 QLs. The solid gray lines represent the DFPT calculation while different symbols for the experimentally determined points correspond to different phonon modes.}
\label{fig:PhononDisp}
\end{figure}

The entire set of data-points obtained from the measurements within the irreducible Brillouin zone is plotted in \autoref{fig:PhononDisp}, superimposed onto the DFPT calculations (grey lines) as obtained for 3 QLs with the inclusion of vdW interactions. Along the $\overline{\mathrm{KM}}$ azimuth there are no data points shown since when scanning along $\overline{\mathrm{\Gamma K M}}$ the intensity of the inelastic features becomes very small for incident angles which are more than approximately $9^{\circ}$ off the specular condition. Hence it is difficult to retrieve spectra with a sufficient signal-to-noise ratio for momentum transfers in the $\overline{\mathrm{\Gamma K M}}$ region.\\
The experimental data points in \autoref{fig:PhononDisp} are plotted as different symbols where each symbol is likely to correspond to a different phonon mode. The majority of the extracted data points corresponds to the RW mode (red crosses) followed by the longitudinal resonance (blue circles) and two further low-lying optical phonon modes (green triangles and orange diamonds). In general the DFPT calculations show excellent agreement with the experimental data points. Note that the inclusion of vdW interactions in the DFPT calculations is particularly important in order to correctly reproduce the longitudinal resonance (blue circles). The role of vdW interactions and SOC will be discussed in greater detail \hyperref[sec:vdWComp]{below}.\\
Interestingly there is no evidence for a strong Kohn anomaly as reported by Howard \emph{et al.}\cite{Howard2013}. According to Howard \emph{et al.} an optical phonon branch originating at 5.8 meV at the $\overline{\Gamma}$-point shows a strong V-shaped indentation at $\Delta K \approx 0.08~\mathrm{\AA}^{-1}$, reaching a minimum of about 4 meV. One reason for the absence of this phonon mode might be the rather low beam energy (8 meV) used in our study. While it is impossible to create phonons with a higher energy than 8 meV, peaks on the annihilation side tend to become weaker with increasing energy. On the other hand, one would expect an indication for this phonon mode for a momentum transfer close to the nesting condition of the Kohn anomaly.\\
Note that the nesting condition $\Delta K = 2 k_F$ for the Kohn anomaly depends on the exact electronic structure, i.e. the position where the Dirac cone crosses the Fermi level $E_F$ which determines $k_F$. In the experiment, the doping of a specific sample plays an important role since the position of the Dirac point with respect to $E_F$ is shifted depending on the doping level. From angle resolved photoemission spectroscopy (ARPES) of the current sample\cite{Tamtogl2017} we know that the Fermi level is 0.35 eV above the Dirac point and the Dirac cone crosses $E_F$ at $k_F = 0.14~\mathrm{\AA}^{-1}$. Hence the nesting condition for the current sample is $\Delta K = 2 k_F = 0.28~\mathrm{\AA}^{-1}$ compared to $\Delta K \approx 0.08~\mathrm{\AA}^{-1}$ in the study by Howard \emph{et al.}\cite{Howard2013}. Moreover, due to the hexagonal warping of the surface state Fermi contour, the nesting should be enhanced when the doping of the sample increases\cite{Fu2009}. Nevertheless there is no evidence for such a Kohn anomaly in our experimental data. In particular since a Kohn anomaly becomes typically more pronounced with decreasing temperature (measurements at 115 K in our study versus room temperature measurements in \cite{Howard2013}). We can only speculate whether the nesting becomes less likely with increasing $\Delta K$, where a spin-flip of the electron is needed, which can be provided by the SO part of the electron-phonon interaction\cite{Chis2013}. Or whether the absence of the Kohn anomaly has to do with some fundamentally different properties of the investigated samples, e.g. the different doping level and possibly different charge carrier densities of the samples.\\
The absence of a Kohn anomaly in our measurements is also in accordance with our DFPT calculations, as shown in \autoref{fig:PhononComp} d) where we report the phonon frequencies at $q_{Nest}=2k_{F}=0.28~\mathrm{\AA}^{-1}$ obtained with a particularly careful sampling of the SBZ as described in the \nameref{sec:CompDetails}. Similar results have been obtained in a recent work by Heid \emph{et al.}\cite{Heid2017} from first principles, who could neither find a large e-ph coupling constant nor the presence of a Kohn anomaly. Hence they suggested that the Kohn anomaly observed in \cite{Howard2013} may be connected to a strong e-ph interaction in the doped bulk material rather than a surface state.\\
Furthermore, a connection with the recently observed acoustic surface plasmon (ASP) on Bi$_2$Se$_3$ is also a possible scenario. The dispersion slope of the ASP on Bi$_2$Se$_3$ was found to be similar to the group velocity of the surface acoustic mode of an Mn doped crystal\cite{Jia2017} as well as the group velocity of the RW mode in our study. Inelastic scattering events associated with such an ASP are likely to appear at a similar energy as the reported Kohn anomaly\cite{Howard2013} making a distinction between different inelastic processes difficult.\\
\begin{table*}[htb]
\centering
\caption{Energy (meV) of the lowest lying surface phonon mode at selected high-symmetry points of Bi$_2$Te$_3$(111) together with a comparison of the theoretically obtained equilibrium geometry of the crystal. The second and third column state whether spin-orbit coupling (SOC) and van der Waals (vdW) interactions where included in the theoretical calculations, respectively. The theoretically obtained values are for films with a thickness of $N$ QLs at a theoretical equilibrium in-plane lattice parameter $a$ (\AA). The average vertical distances $\Delta z$ between different atoms are also reported. Here Te2 indicates the outermost Te atom in a QL and Te1 the central one (see \autoref{fig:struct}(a)). The in-plane atomic positions are fixed due to symmetry.}
\begin{minipage}{\textwidth}\centering
\begin{tabular}{ l c c c . . . . . . }
\toprule
\multicolumn{1}{c }{Method} &  \multicolumn{1}{c }{SOC} &  \multicolumn{1}{c }{vdW} & \multicolumn{1}{c }{$N$} & \multicolumn{1}{c }{$a$ (\AA)} & \multicolumn{1}{c }{$\Delta z$ Te2-Te2 (\AA)} & \multicolumn{1}{c }{$\Delta z$ Te2-Bi (\AA)} & \multicolumn{1}{c }{$\Delta z$ Te1-Bi (\AA)}  & \multicolumn{1}{c }{$\overline{\mathrm{ M }}$} & \multicolumn{1}{c }{$\overline{\mathrm{ K }}$} \\ \addlinespace
\midrule
Theory (DFPT)\cite{Chis2012} & no & no & 1 & 4.40 & & & & 4.5 & 5.6  \\
Theory (vdW-DF2-C09)  \footnote{Value obtained from current study\label{note1}} & no & yes & 1 & 4.389 & $-$ & 1.730 & 2.049 & 3.70 & 4.78  \\
Theory (vdW-DF2-C09)  \textsuperscript{\ref{note1}} & no & yes & 2 & 4.389 & 2.688 & 1.740 & 2.037 & 3.79 & 4.79  \\
Theory (vdW-DF2-C09) \textsuperscript{\ref{note1}} & no & yes & 3 & 4.389 & 2.681 & 1.736 & 2.036 & 3.9 & 4.80  \\
Theory (vdW-DF2-C09) \textsuperscript{\ref{note1}} & no & yes & 5 & 4.389 & 2.682 & 1.736 & 2.036 & 3.9 & 4.80  \\
Theory (PBE+D3)  \textsuperscript{\ref{note1}} & no & yes & 3 & 4.324 & 2.868 & 1.761 & 2.049 & 3.59 & 4.76  \\
Theory (PBE+D3) \textsuperscript{\ref{note1}} & yes & yes & 3 & 4.333 & 2.736 & 1.780 & 2.061 & 3.08 & 4.12  \\
Theory (PBE) \textsuperscript{\ref{note1}} & yes & no & 3 & 4.468 & 2.759 & 1.744 & 2.063 & 3.39 & 4.40  \\
Theory (PBE) \textsuperscript{\ref{note1}} & no & no & 3 & 4.446 & 3.091 & 1.723 & 2.055 & 3.81 & 4.96  \\
Theory (PBE) \cite{Huang2012} & yes & no & 5 & 4.386  \footnote{fixed at experimental value\label{note2}} & & & & 2.95 & 4.00  \\
\cmidrule{1-10}
Experiment (HAS) \textsuperscript{\ref{note1}} & - & - & - & 4.36 & $-$  & $-$  & $-$ & 3.7 & 4.7  \\  
Experiment (X-ray) \cite{Nakajim1963} & - & - & - & 4.386 & 2.613  & 1.743  & 2.033 &  &   \\  
\bottomrule
\end{tabular}
\label{tab:LambdaTable}
\end{minipage}
\end{table*}
Finally, upon close inspection there appears to be a small indentation for the L-branch at $\Delta K \approx 0.35~\mathrm{\AA}^{-1}$ along $\overline{\Gamma \mathrm{M}}$. The evidence is not particularly strong based on the number of data points in \autoref{fig:PhononDisp} but is better seen in the intensity plot (see the red arrow in \autoref{fig:3D_phonon} at $0.35~\mathrm{\AA}^{-1}$). Since in the present Bi$_2$Te$_3$ sample the Fermi level is above the bulk conduction band minimum\cite{Tamtogl2017}, the impinging $^3$He atom detects some mixing of the Dirac states with bulk states, thus relaxing the spin selection rule and making such a transition possible. As the Dirac states are 2D they do not depend on $k_z$, but their ARPES intensity strongly depends on $k_z$, depending on the Fourier components of the Dirac wavefunction in the $z$-direction\cite{Tamtogl2017,Michiardi2014}. As can be seen from ARPES data of the current sample (Fig. 4 of Ref.\cite{Tamtogl2017}), the Fermi level is above the minimum of the conduction band and the minimum in the $k_z$ direction occurs at $k_z = 0.31~\mathrm{\AA}^{-1}$, where the Dirac cone intersects the Fermi level at $2 k_F \approx 0.35~\mathrm{\AA}^{-1}$. Thus the intensification of the Dirac state at the Fermi level is probably due to the resonance (mixing) of this state with bulk states. If this is the case the spin selection rule is considerably relaxed, which explains why the electron-hole excitation is detected by HeSE, though it is rather weak.

\subsection{The role of vdW interactions and SOC}
\label{sec:vdWComp}
After this ample discussion about possible anomalies in the vibrational modes of Bi$_2$Te$_3$ we will now address the question to what degree DFPT calculations reproduce specific phonon energies. Therefore, we performed calculations including different vdW corrections and, where possible, both including and neglecting SOC.\\
Firstly, we start with a short discussion regarding the equilibrium lattice parameters as obtained with different theoretical methods since changes in the equilibrium geometry will lead to different phonon spectra. \autoref{tab:LambdaTable} shows the obtained theoretical equilibrium in-plane lattice parameter $a$ as well as the vertical distances $\Delta z$ between different atoms (see \autoref{fig:struct}(a)) for different functionals and films with a thickness of $N$ QLs. In terms of the in-plane lattice parameter $a$, all functionals perform well and agree with the experimentally determined parameters up to the second decimal place, except for the PBE functional without dispersion corrections which gives a slightly larger $a$. The vertical distances within one QL are reproduced very well by all functionals. However, the vdW gap between the QLs (Te2-Te2 distance in \autoref{tab:LambdaTable}) is described most accurately with the vdW-DF2-C09 functional while PBE tends to give larger distances which becomes at most 0.5 \AA\ larger than the experimental one if neither SOC nor vdW are accounted for. Hence we can already anticipate that the vdW-DF2-C09 functional, which gives the best agreement for the vdW gap, will also provide a good description of the low-energy acoustic phonon modes. Moreover, we have also performed calculations with the vdW-DF2-C09 functional for a number of different QLs $N = 1, 2, 3$ and $5$, which are shown in \autoref{tab:LambdaTable}. It can be seen that changes on the equilibrium geometry are already minor when going from 2 to 3 QLs while there are no significant differences when going from 3 to 5 QLs, suggesting that a DFPT calculation for 3 QLs is sufficient for an accurate mapping of the entire phonon dispersion relation.\\
\begin{figure*}[tb]
\centering
\includegraphics[width=0.65\textwidth]{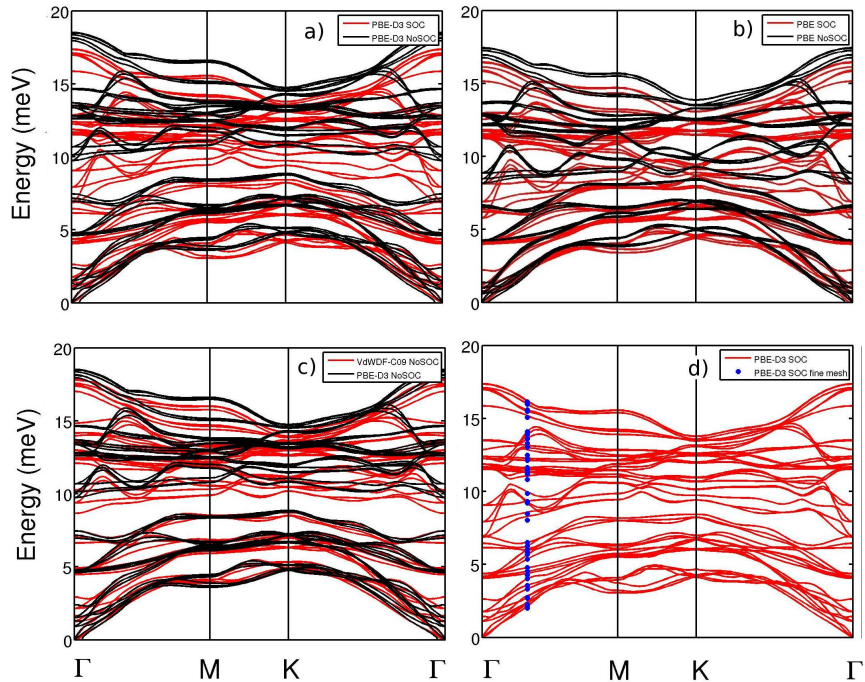}
\caption{(Colour online) a) Phonon dispersions computed using PBE with the inclusion of the vdW PBE-D3 correction including (red) and neglecting (black) SOC. b) Phonon dispersion computed using bare PBE including (red) and neglecting (black) SOC. c) Comparison between phonon dispersion computed with two different vdW corrections PBE-D3 (red) and non-local functional vdW-DF2-C09 (black) without SOC. d) Phonon dispersion computed with PBE-D3 with SOC on a standard mesh (red lines) and phonon frequencies computed at a $q$-point corresponding to the nesting vector of the Fermi surface (blue dots) for a system slightly doped in order to reproduce the experimental position of the Dirac cone. The calculation has been carried out with a very fine grid around the $\Gamma$-point and a small smearing in order to allow a proper description of a potential Kohn anomaly.}
\label{fig:PhononComp}
\end{figure*} 
\autoref{tab:LambdaTable} shows a comparison for the experimentally determined energy of the lowest lying surface phonon mode at the $\overline{\mathrm{M}}$ and $\overline{\mathrm{K}}$-point with several calculations. Note that the experimental value for the $\overline{\mathrm{K}}$ point has been extrapolated using a sine fit to the experimental points shown in \autoref{fig:PhononDisp}. Our results show generally a good agreement with the experimentally measured lowest lying surface phonon mode except for the PBE functional when including both vdW corrections and SOC which tends to give rise to softer phonon energies at the Brillouin zone boundary. \autoref{tab:LambdaTable} reveals a rather feeble dependence with respect to vdW correction or SOC. The comparison shows that DFPT (without SOC, considering only 1 QL\cite{Chis2012}) tends to overestimate the vibrational energies. On the other hand Huang\cite{Huang2012} uses up to 5 QLs for DFPT calculations based on the PBE approximation with SOC but obtains smaller energies than compared to the experiment. However, in Ref.\cite{Huang2012} the in-plane lattice parameter was fixed at the experimental one. Based on the above described calculations with increasing number of QLs it is more likely that the reported soft phonon energies for 5 QLs\cite{Huang2012} are a consequence of the fixed lattice parameter rather than the number of QLs. Indeed we see from our DFPT calculations with the vdW-DF2-C09 functional for $N = 1, 2, 3$ and $5$ in \autoref{tab:LambdaTable} that the influence of the number of QLs on the phonon energies at the Brillouin zone boundaries ($\overline{\mathrm{M}}$,$\overline{\mathrm{K}}$) is very weak. The reason can be found in the weak vdW coupling between the layers, so that particularly at the zone boundaries where only short-range interactions are important, the phonons of a single QL resemble closely the ones of multiple slabs. On the other hand the vdW interaction is crucial in order to obtain the proper geometry as stated above.\\
In terms of the lattice vibrations, vdW interactions are instead much more important close to $\overline{\Gamma}$ where long-range interactions are important for the phonon frequencies. Here two effects of the vdW interaction play an important role: The first one is again for obtaining the right equilibrium geometry, the second one is the direct contribution of the vdW interactions to the force constants. This is particularly important in order to obtain the correct longitudinal resonance (blue circles of the experimental data, with about 2 meV at $\overline{\Gamma}$) which is further described below.\\
The influence of vdW corrections and SOC on the phonon energies can be best seen by a direct comparison of the phonon dispersion obtained with different theoretical methods (for 3 QLs) as plotted in \autoref{fig:PhononComp}. First of all, we note that the agreement with experiments is marginally better without the inclusion of SOC, likely due to a compensation of errors between the underbinding problem that often characterises the PBE functional and the contribution of SOC. The effect of SOC is generally weak for all low energy modes while it becomes increasingly important for the high-energy optical phonons where it is responsible for a softening of up to 13\% (See \autoref{fig:PhononComp} a) and b)).\\
\begin{figure*}[ht]
\centering
\includegraphics[width=0.8\textwidth]{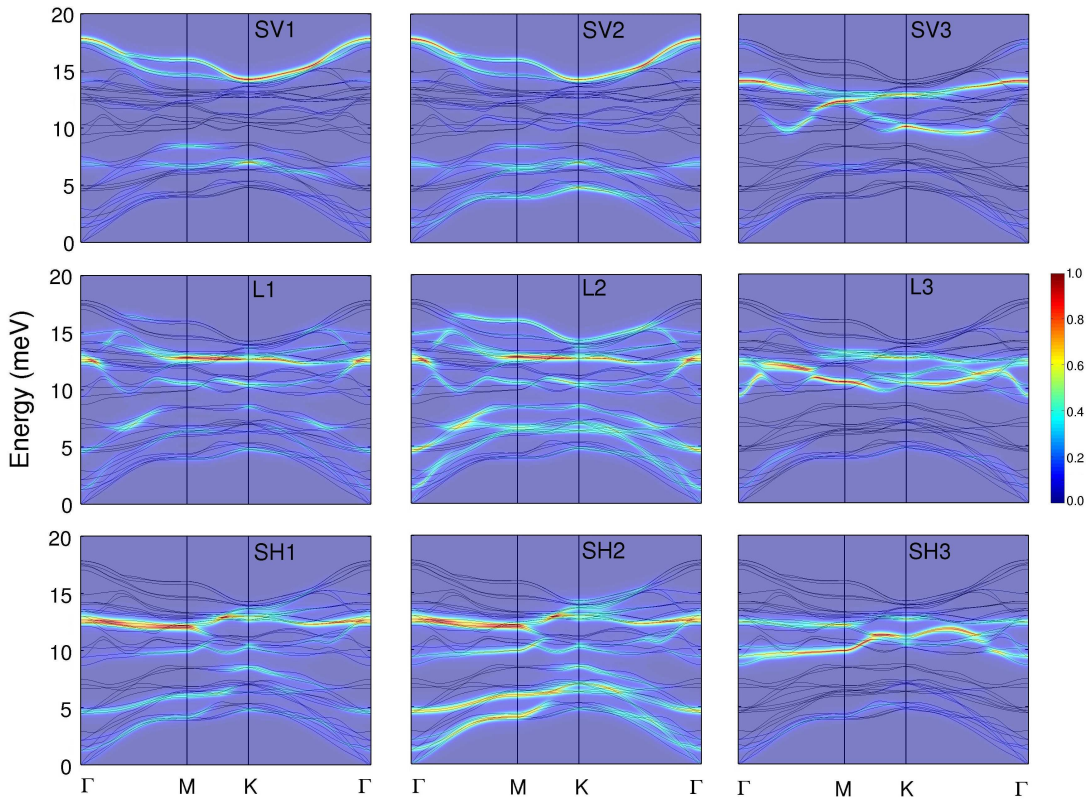}
\caption{(Colour online) Polarisation of the calculated phonon dispersion. The intensity corresponds to shear vertical (SV), longitudinal (L) and shear horizontal (SH) polarisation projected onto the first or top-most atomic layer (1), second layer (2) and third layer (3), respectively.}
\label{fig:PhononProj}
\end{figure*} 
As might be expected, vdW corrections are instead crucial for a good description of the low energy optical modes (particularly the ones starting at around 2 and 5 meV at $\Gamma$) involving a modulation of the inter-QL distance as can be inferred by comparing \autoref{fig:PhononComp}b) reporting the results of bare PBE without vdW corrections with \autoref{fig:PhononComp}a) and c) in which two different types of vdW corrections are considered. In particular, the non-local exchange-correlation functional vdW-DF2-C09 appears to be the most accurate in describing the weak coupling between the QLs since it is the only one that reproduces the experimental mode at 2 meV (see \autoref{fig:PhononDisp}).

\subsection{Polarisation of the surface phonon modes}
Finally, we turn to the polarisation analysis of the surface phonon modes. The full set of calculated dispersion relations along the symmetry directions is shown in \autoref{fig:PhononProj} together with the corresponding phonon densities projected onto the first, second and third layer for shear vertical (SV1, SV2, SV3), longitudinal (L1, L2, L3) and shear horizontal (SH1, SH2, SH3) polarisations. Since the optical modes inside the QLs are weakly affected by the surface, at $\Delta K = 0$ they retain an approximate gerade or ungerade symmetry with respect to the central Te layer. They can then be easily recognised from their third-layer intensity (\autoref{fig:PhononProj}: SV3, L3, SH3), which is comparatively large for quasi-ungerade and almost vanishing for quasi-gerade modes.\\
Based on the DFPT calculations, the polarisation of the experimentally observed, particularly intense, RW mode is a combination of shear vertical and shear horizontal, which is however only possible along $\overline{\Gamma \mathrm{K}}$. Interestingly the largest intensity comes from the SH2 mode (central panel at the bottom of \autoref{fig:PhononProj}). Note that if the scattering plane, defined by the incoming and scattered He beam, coincides with a mirror plane of the surface, the detection of purely SH modes is in principle forbidden due to symmetry reasons\cite{Tamtogl2015}. However, while the atomic displacement is largest for the SH2 polarisation, we see from \autoref{fig:PhononProj} that the mode always retains an SV polarisation component as well. Moreover, even a purely SH mode may give rise to CDOs above the first atomic layer which are eventually observed in inelastic He atom scattering.\\
The phonon mode at 2 meV is likely ascribable to a shear vertical mode in which the entire outermost QL vibrates rigidly against the inner ones at the $\Gamma$-point which then enters a longitudinal resonance when moving towards the zone border. The mode at 5 meV is instead related mainly to a longitudinal vibration of the first and second atomic layer moving in anti-phase with the fourth and fifth layer while the 8 meV mode is characterised by a shear vertical vibration involving the same pair of atoms with a similar anti-phase pattern.

\section{Summary and conclusion}
Due to being very low in energy the acoustic phonon modes on topological insulator surfaces are difficult to address and resolve experimentally. However, a precise knowledge of the surface dynamical properties is one of the key ingredients to understand fundamental properties of archetypal topological insulators at elevated temperatures (where electron-phonon scattering processes may become important) as well as for the use of Bi$_2$Te$_3$ as one of the most efficient thermoelectric materials in nanoscale devices\cite{Liang2016,Rojo2017}.\\
We have measured the surface lattice vibrations of Bi$_2$Te$_3$(111) in the low energy region using high-resolution helium-3 spin-echo spectroscopy. The low energy region of the surface phonons is mainly dominated by the Rayleigh wave which has been found to be absent in previous experimental studies. The appearance of the RW is consistent with previous bulk lattice dynamics studies as well as theoretical predictions of the surface phonon modes. The speed of sound of the RW is between $1300~\mbox{m}/\mbox{s}$ and $1600~\mbox{m}/\mbox{s}$ along the $\overline{\mathrm{\Gamma M}}$ and $\overline{\mathrm{\Gamma K}}$ directions, respectively. Furthermore, our results do not support the presence of a Kohn anomaly, connected with a surface phonon mode, as inferred from a previous study.\\ 
The surface phonon dispersion for 3 quintuple layers has been calculated with vdW corrected DFPT calculations (without SOC). Comparison between the experimentally determined values and the calculations show excellent agreement. We have also compared the experimentally determined phonon energies at a number of high symmetry points with calculations using different functionals for the treatment of SOC and vdW interactions. Our results suggest that in order to calculate the surface lattice vibrations of three-dimensional topological insulators with their layered structure, the inclusion of vdW interactions is essential. Within the methods used in our study, the non-local exchange-correlation functional vdW-DF2-C09 appears to be the most accurate one in describing the coupling between the layers since it provides the right distance for the vdW gap and is the only method that correctly reproduces the experimentally observed longitudinal resonance.\\
While the performance of the vdW-DF2-C09 functional in terms of an accurate description of the lattice vibrations of Bi$_2$Te$_3$ is excellent, it should be noted that one shortcoming is the currently missing implementation of SOC. For now, it may be advocated for further first-principles vibrational investigations of layered TI materials while other methods may be more appropriate when it comes to calculations of the electronic structure, in particular, if they involve non-adiabatic processes within the context of electron-phonon interaction. Future efforts towards a correct description of both vdW interactions as well as SOC may provide significant insight into such complex mechanisms in TI materials.\\ 
The observation of a prominent surface acoustic may have important implications in particular for thin-film and nanoscale devices: The thermal conductivity of such devices may be strongly influenced by the contribution of the observed acoustic mode. Moreover, the improved accuracy when including vdW interactions shows that these are possibly crucial for an exact theoretical description of application-relevant issues like the thermal conductivity of layered structures in general.

\section*{Acknowledgment}
One of us (A.T.) acknowledges financial support provided by the FWF (Austrian Science Fund) within the projects J3479-N20 and P29641-N36. The authors are grateful for financial support by the Aarhus University Research Foundation, VILLUM FOUNDATION via the Centre of Excellence for Dirac Materials (Grant No. 11744) and the SPP1666 of the DFG (Grant No. HO 5150/1-2). M.B., E.M.J.H. and B.B.I. acknowledge financial support from the Center of Materials Crystallography (CMC) and the Danish National Research Foundation (DNRF93).

\bibliography{literature}

\end{document}